\documentclass[conference]{IEEEtran}
\usepackage[letterpaper, left=0.65in, right=0.65in, bottom=1.02in, top=0.75in]{geometry}
\IEEEoverridecommandlockouts

\usepackage{amsmath,amssymb,amsfonts}
\usepackage{algorithmic}
\usepackage{algorithm}
\usepackage{graphicx}
\usepackage{textcomp}
\usepackage{cite}
\usepackage{bm}
\usepackage{color}

\newcommand{\bh}{\hat{\mathbf{h}}}
\newcommand{\bH}{\mathbf{H}}
\newcommand{\ba}{\mathbf{a}}
\newcommand{\bd}{\mathbf{d}}
\newcommand{\bs}{\mathbf{s}}

\newcommand{\bW}{\mathbf{W}}
\newcommand{\bu}{\mathbf{u}}
\newcommand{\by}{\mathbf{y}}
\newcommand{\bn}{\mathbf{n}}
\newcommand{\bI}{\mathbf{I}}

\begin{document}

\title{Integrated Sensing and Communications over\\
Hierarchical Cellular and Cell-Free MIMO Systems\\
\thanks{This work was supported by the German Federal Ministry of Education and Research (BMBF) through \emph{Open6GHub+} (Grant no.  \emph{16KIS2402K}) project.}
}

\author{
\IEEEauthorblockN{Wei Jiang}
\IEEEauthorblockA{Intelligent Networking Research Group\\
German Research Center for Artificial Intelligence (DFKI)\\
Kaiserslautern, 67663 Germany}
\and
\IEEEauthorblockN{Hans D. Schotten}
\IEEEauthorblockA{Department of Electrical and Computer Engineering\\
University of Kaiserslautern (RPTU)\\
Kaiserslautern, 67663 Germany}
}

\maketitle

\begin{abstract}
This paper studies integrated sensing and communications (ISAC) over a hybrid system that seamlessly combines legacy cellular base stations with distributed cell-free (CF) access points (APs). We propose a hierarchical ISAC architecture where a central base station (CBS) serves its near users and simultaneously operates as a monostatic radar for aerial target detection, while distributed APs---many idle under user-centric clustering---act as cost-free bistatic receivers. The CBS jointly handles communication processing and multi-static sensing fusion, reducing fronthaul overhead compared to conventional cell-free ISAC. To achieve this, a five-phase time-division duplexing workflow with precise ISAC role assignment is specified. Closed-form expressions for spectral efficiency and multi-static sensing signal-to-noise ratio analytically characterize the communications--sensing Pareto frontier. Numerical results confirm that the proposed hierarchical design simultaneously achieves higher sum throughput and superior sensing accuracy than conventional cell-free ISAC.
\end{abstract}


\section{Introduction}
\label{sec:intro}

\IEEEPARstart{C}{ell-free} (CF) massive multi-input multi-output (MIMO) \cite{Ngo2017} has emerged as a promising sixth-generation (6G) technology \cite{Jiang2021}, where a large number of distributed access points (APs) jointly serve all users, delivering uniform coverage that conventional cellular networks cannot match \cite{Jiang2023EUCNC}. Integrated sensing and communications (ISAC) \cite{Liu2022JSAC} is another 6G pillar, enabling a single waveform and shared hardware platform to simultaneously perform wireless communication and radar sensing \cite{Hua2023TVT}. Deploying ISAC over a CF network (we call it CF-ISAC hereinafter) is therefore a promising direction, as the spatially distributed APs provide not only angular diversity for sensing but also enhanced communications coverage.

Despite its potential, the CF architecture faces significant practical barriers. Acquiring hundreds of suitable AP sites is costly and often infeasible, and the associated fronthaul network demands labour-intensive civil works such as trenching and fibre laying, substantially increasing deployment time and cost \cite{Kim2022}. To address this, we proposed a hierarchical cell-free (HCF) architecture in \cite{Jiang2024ICC}, which reuses a legacy cellular base station (BS) at the coverage center alongside a reduced scale of distributed APs, and extended the concept to wideband scenarios in \cite{Jiang2024GC}. A similar hierarchical framework was independently discussed in \cite{Zhang2024}, and recent studies have explored integrating CF massive MIMO into legacy cellular infrastructure through cooperation mechanisms \cite{Buzzi2024}.

CF-ISAC has attracted growing research interest. Demirhan and Alkhateeb \cite{Demirhan2024} derived a multi-static sensing signal-to-noise ratio (SNR) and proposed joint sensing and communication beamforming design. Behdad \textit{et al.} \cite{Behdad2024} studied power allocation for simultaneous communication and target detection. Liu \textit{et al.} \cite{Liu2024} and Elfiatoure \textit{et al.} \cite{Elfiatoure2024} investigated AP mode selection, assigning each node as either a transmitter or a passive echo receiver. Mao \textit{et al.} \cite{Mao2024} characterised the communications--sensing trade-off region, accounting for pilot contamination caused by target reflections. Ren \textit{et al.} \cite{Ren2024} addressed secure CF-ISAC against both communication and sensing eavesdroppers. However, none of these works considers legacy BSs from an existing cellular network or exploits the cooperation between a BS and distributed APs.

This paper extends ISAC to hierarchical CF (hereafter called HCF-ISAC) and makes the following contributions.
\begin{itemize}
\item We identify three properties that render HCF-ISAC advantageous over CF-ISAC: (i) the BS provides co-located radar antenna aperture with full coherent array gain; (ii) APs that are idle under user-centric (UC) clustering are repurposed as bistatic echo receivers at zero hardware cost; and (iii) the BS simultaneously serves as the communications processor and the sensing fusion center.
\item \textit{A novel HCF-ISAC protocol.} We assign ISAC roles to each node type---BS, active AP, and idle AP---and present a five-phase workflow that integrates communications and sensing within a single coherent interval, consistent with the CF-ISAC frameworks of \cite{Demirhan2024,Elfiatoure2024}.
\item  We derive spectral efficiency (SE) expressions that include a sensing-interference term, and a multi-static sensing SNR separating the monostatic contribution from the bistatic sum, together yielding a closed-form characterisation of the communications--sensing Pareto frontier.
\end{itemize}

The remainder of this paper is organized as follows. Section~\ref{sec:system} presents the system model. Section~\ref{sec:workflow} assigns node roles and specifies the HCF-ISAC protocol. Section~\ref{sec:analysis} provides the performance analysis. Section~\ref{sec:results} presents numerical results. Section~\ref{sec:conclusion} concludes the paper.

\begin{figure}[!t]
  \centering
  \includegraphics[width=0.85\columnwidth]{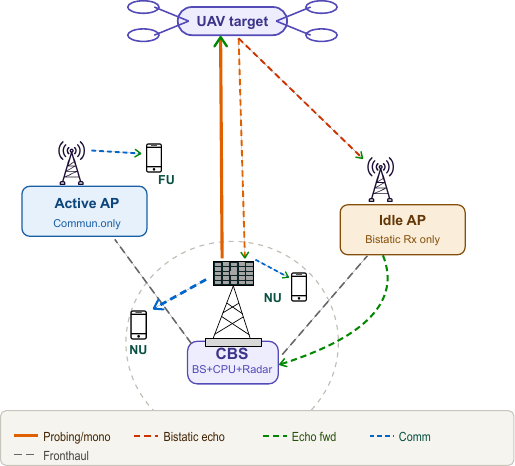}
  \caption{Illustration of the HCF-ISAC architecture.}
  \label{systemModelISACHCF}
\end{figure}

\section{System Model}
\label{sec:system}

\subsection{Hierarchical Network Topology}

We consider the hierarchical system of \cite{Jiang2024ICC}, where a central BS (CBS) with $N_b$ antennas is located near the center of a coverage area, as  illustrated in \figurename~\ref{systemModelISACHCF}. A total of $M$ antennas serve $K \ll M$ user equipments (UEs) in time-division duplexing (TDD) mode.  CBS also functions as the central processing unit (CPU) of the $M - N_b$ single-antenna distributed APs, which are deployed surrounded the CBS, especially its edge area, and are connected via a fronthaul network. CBS classifies each UE~$k$ as a \emph{near user} (NU) or \emph{far user} (FU), according to
\begin{equation}
    \label{eq:nu_set} \mathcal{K}_0 = \{k : \beta^{0}_{k} \geq \bar{\beta}^{0}\},
\end{equation}
where $\beta^{0}_{k}$ is the CBS-to-UE~$k$ large-scale fading coefficient and $\bar{\beta}^{0}$ is a CBS-specific threshold. For each FU $k \notin \mathcal{K}_0$, the CBS forms an UC cluster of APs as $\mathcal{M}_k = \{m : \beta_{mk} \geq \bar{\beta}_{k}\}$, where $\beta_{mk}$ denotes the large-scale fading between AP~$m$ and UE~$k$, and $\bar{\beta}_{k} = \frac{1}{M - N_b}\sum_{m} \beta_{mk}$ is the per-user threshold from \cite{Jiang2024ICC}, although it can be optimized based on other criteria. The set of \emph{active} APs is $\mathcal{M}_{\text{act}} = \{m : \mathcal{K}_m \neq \emptyset\}$, where $\mathcal{K}_m = \{k : m \in \mathcal{M}_k\}$ denotes the set of users served by AP $m$, and the set of \emph{idle} APs is $\mathcal{M}_{\text{idle}} = \{m : \mathcal{K}_m = \emptyset\}$.

\subsection{Communications Channel Model}

The scalar channel between AP~$m$ and UE~$k$ is denoted by $g_{mk} \sim \mathcal{CN}(0, \beta_{mk})$, where $\beta_{mk}$ embeds COST-Hata path loss and log-normal shadowing \cite{Ngo2017}. The $N_b \times 1$ channel from the CBS to UE~$k$ is $\mathbf{h}_k \sim \mathcal{CN}(\mathbf{0}, \beta^{0}_{k}\bI_{N_b})$. Linear minimum mean-square-error (MMSE) estimates are $\bh_k \sim \mathcal{CN}(\mathbf{0}, \alpha^{0}_{k}\bI_{N_b})$ with $\alpha^{0}_{k} = \frac{p_u (\beta^{0}_{k})^2}{p_u \beta^{0}_{k} + \sigma^{2}_{n}}$, and $\hat{g}_{mk} \sim \mathcal{CN}(0, \alpha_{mk})$ with $\alpha_{mk} = p_u \beta^{2}_{mk}/(p_u \beta_{mk} + \sigma^{2}_{n})$, where $p_u$ is the UE transmit power and $\sigma^{2}_{n}$ is the noise variance.

\subsection{Sensing Channel Model}

The system detects a single unmanned aerial vehicle (UAV) target located at distance $d_t$,
azimuth $\theta_t$, and elevation $\phi_t$ relative to the CBS.
Due to the UAV's high altitude, all target-related links are modelled as
line-of-sight (LoS)~\cite{Behdad2024}. 
The channel vector from the CBS to the UAV is $\mathbf{h}_{bt} = \sqrt{\zeta_{bt}}\,\ba(\theta_t,\phi_t)$, where $\zeta_{bt}$ is the one-way path loss from the CBS to the UAV, and $\ba(\theta_t,\phi_t)\in\mathbb{C}^{N_b}$ is the CBS steering vector.
Assuming a uniform linear array (ULA) with half-wavelength spacing,
\begin{equation}\label{eq:steering}
    \ba(\theta,\phi) = \frac{1}{\sqrt{N_b}}
    \Bigl[1,\,e^{j\pi\sin\phi\cos\theta},\,\ldots,\,
    e^{j\pi(N_b-1)\sin\phi\cos\theta}\Bigr]^{\!T}.
\end{equation}
Each AP~$m$ is equipped with a single antenna, so the LoS channel
from AP~$m$ to the UAV reduces to the scalar $h_{mt} = \sqrt{\zeta_{mt}}\,e^{j\psi_{mt}}$, where $\zeta_{mt}$ is the one-way path loss from the UAV to AP~$m$,
and $\psi_{mt}$ is the associated LoS phase offset\footnote{The 
absolute LoS phase $e^{j\psi_{bt}}$ is omitted as it is a common 
scalar across all array elements and is absorbed into $\xi_t$ without 
loss of generality.}.

The monostatic CBS\,$\to$\,UAV\,$\to$\,CBS reflected channel is given by
\begin{equation}\label{eq:mono_ch}
    \bH_{\mathrm{mono}} = \xi_t\,e^{j2\pi f_{D,bt}\,t}\,\mathbf{h}_{bt}\mathbf{h}_{bt}^{T},
\end{equation}
where $\xi_t \in \mathbb{C}$ is the reflection coefficient determined
by the UAV's radar cross section (RCS), and the Doppler shift $f_{D,bt} = \frac{2v_t}{\lambda}\cos(\varphi_t - \theta_t)$. Here, $v_t$ is the UAV speed, $\varphi_t$ is its flying direction, and $\lambda = c/f_c$ is the carrier wavelength. Moverover, the CBS\,$\to$\,UAV\,$\to$\,AP~$m$ bistatic channel is
\begin{equation}\label{eq:bi_ch}
    \mathbf{c}_{mt} = \xi_t\,e^{j2\pi f_{D,mt}\,t}\,h_{mt}\,\mathbf{h}_{bt}^{T}
    \in \mathbb{C}^{1\times N_b},
\end{equation}
where $f_{D,mt} = \frac{v_t}{\lambda}
    \bigl[\cos(\varphi_t-\theta_t)+\cos(\varphi_t-\theta_{mt})\bigr]$ with $\theta_{mt}$ the angle of departure from AP~$m$ to the UAV.

\section{The proposed HCF-ISAC Protocol}
\label{sec:workflow}
\subsection{Node Role Assignment}
Inspired by the AP mode-selection concept of \cite{Liu2024,Elfiatoure2024}, we assign ISAC roles implicitly from the UC clustering outcome, eliminating any additional optimization step for role determination\footnote{The three-way role split parallels the C-AP\,/\,S-AP paradigm of \cite{Elfiatoure2024}, but in HCF no separate optimisation is needed: the network structure determines roles deterministically. Three structural properties set HCF apart from the CF networks studied in \cite{Demirhan2024,Behdad2024,Liu2024,Elfiatoure2024,Mao2024,Ren2024}.
\textit{Property~1---Co-located monostatic aperture:} The CBS provides coherent $N_b$-element monostatic radar processing; the monostatic sensing SNR scales as $N_b^2$ (two-way array gain), which no individual distributed AP in a uniform CF network can match.
\textit{Property~2---Free bistatic receivers:} Under UC clustering, the fraction of idle APs $|\mathcal{M}_{\text{idle}}|/(M - N_b)$ grows as $K/M$ decreases (the typical CF regime). These nodes extend the sensing aperture at zero cost; in conventional CF-ISAC no idle APs exist by current design.
\textit{Property~3---Unified fusion centre:} Because the CBS is already the CPU, sensing fusion and communications detection share the same hardware. The only new fronthaul data are the scalar echo samples from idle APs, whereas \cite{Demirhan2024,Ren2024} require full inter-AP coordination.}. As illustrated in \figurename~\ref{systemModelISACHCF}, 
\begin{itemize}
    \item \textit{CBS---Four Simultaneous Functions}: \begin{enumerate}
\item \emph{Communication hub for NUs}: transmits to all $k \in \mathcal{K}_0$ using downlink precoding with per-antenna power $p_d$.
\item \emph{CPU}: coordinates active APs for FU communications via fronthaul, exactly as in \cite{Jiang2024ICC}.
\item \emph{Monostatic radar transmitter}: emits a dedicated probing beam $\bs_r \in \mathbb{C}^{N_b}$ steered toward the UAV.
\item \emph{Sensing fusion centre}: processes its own monostatic echo and aggregates bistatic echo scalars from all sensing APs.
\end{enumerate}
\item \textit{Active APs}: An active AP $m \in \mathcal{M}_{\text{act}}$ sends data toward its FU cluster $\mathcal{K}_m$ during the downlink. In the uplink, it receives FU signals and forwards the scalar statistics $\hat{g}^{*}_{mk} y_m$, $k \in \mathcal{K}_m$, to the CBS. 
\item \textit{Idle APs}: An idle AP $m \in \mathcal{M}_{\text{idle}}$ has no communications function. In a pure communications system it would be switched off. In HCF-ISAC it is activated in passive receive mode to collect the bistatic echo $r_m$ of the CBS probing beam reflected off the UAV. Each idle AP forwards only the scalar $r_m$ to the CBS, incurring negligible additional fronthaul load compared to the multi-symbol statistics exchanged by active APs. 
\end{itemize}

\subsection{Five-Phase TDD Workflow}

\begin{algorithm}[t]
\caption{Five-Phase TDD Protocol for HCF-ISAC}
\label{alg:tdd}
\begin{algorithmic}[1]
\REQUIRE $\tau_p$ pilot symbols, $\rho \in [0,1]$, $p_d$, $p_u$, $(\hat{\theta}^{(0)}_t, \hat{\phi}^{(0)}_t)$
\ENSURE $(\hat{\theta}^{(n)}_t, \hat{\phi}^{(n)}_t, \hat{d}^{(n)}_t)$ for each frame~$n$ 
\FOR{$n = 1, 2, \ldots$}
\STATE \textit{UL training:} UEs transmit $\tau_p$-length pilots with power $p_u$. CBS estimates $\bh_k$, $\forall k$; classifies $k \in \mathcal{K}_0$ if $\beta^{0}_{k} \geq \bar{\beta}^{0}$. AP~$m$ estimates $\hat{g}_{mk}$; sets $\mathcal{M}_{\text{idle}} = \{m : \mathcal{K}_m = \emptyset\}$.
\STATE \textit{Beamforming design:} CBS forms precoder $\bW_c$ and radar beam $\bs^{(n)}_r = \sqrt{N_bp_d}\, \ba(\hat{\theta}^{(n-1)}_t, \hat{\phi}^{(n-1)}_t)$; active AP~$m$ sets $w_{mk} = \hat{g}^{*}_{mk}/\sqrt{\mathbb{E}[|\hat{g}_{mk}|^2]}$.
\STATE \textit{DL and sensing probe:} CBS sends
$\bd = \sqrt{1-\rho}\,\bW_c \bu + \sqrt{\rho}\,\bs^{(n)}_r$; active AP~$m$ transmits $s_m$.
\STATE \textit{Echo reception:} CBS collects monostatic echo $\by_b$; AP~$m\in \mathcal{M}_{\text{sense}}$ collects and forwards $r_m$ to CBS.
\STATE \textit{Parameter estimation:} CBS computes $z_b = \ba^H(\hat{\theta}_t, \hat{\phi}_t)\by_b$, $z_{\text{bi}} = \sum_{m \in \mathcal{M}_{\text{sense}}} \sqrt{\xi_{mt}}\, e^{-j\psi_{mt}} r_m$, and fuses $\{z_b, z_{\text{bi}}\}$ via MUSIC/ESPRIT to obtain $(\hat{\theta}^{(n)}_t, \hat{\phi}^{(n)}_t, \hat{d}^{(n)}_t)$; updates $\bs^{(n+1)}_r \leftarrow \sqrt{p_d}\, \ba(\hat{\theta}^{(n)}_t, \hat{\phi}^{(n)}_t)$.
\ENDFOR
\end{algorithmic}
\end{algorithm}

Under block fading, the coherent interval is divided into five phases, as summarized in Algorithm~\ref{alg:tdd}. That is:
\subsubsection{Uplink Training}
All $K$ UEs transmit mutually orthogonal pilots of length $\tau_p \geq K$. The CBS obtains $\bh_k$ for all $k$ and each AP~$m$ obtains $\{\hat{g}_{mk}\}$ via MMSE estimation. The CBS then classifies UEs per (\ref{eq:nu_set}), constructs clusters $\{\mathcal{M}_k\}$, and identifies $\mathcal{M}_{\text{idle}}$. Pilot contamination due to UAV reflections \cite{Mao2024} is negligible for $\tau_p \geq K$.

\subsubsection{Beamforming Design}
The CBS allocates its total power $N_b p_d$ between communications 
and sensing via a power-splitting factor $\rho \in [0,1]$. The CBS 
downlink signal is
\begin{equation}\label{eq:cbs_dl}
    \bd = \sqrt{1-\rho}\,\bW_c \bu_{\mathcal{K}_0} + \sqrt{\rho}\,\bs_r,
\end{equation}
where $\bu_{\mathcal{K}_0}$ is the stacked NU data vector with 
$\mathbb{E}[\bu\bu^H] = \bI$, $\bs_r$ is the probing vector, and 
$\bW_c \in \mathbb{C}^{N_b \times |\mathcal{K}_0|}$ is the precoding 
matrix. Applying conjugate beamforming (CBF), we have\footnote{The 
proposed hierarchical ISAC framework is precoding-agnostic and 
accommodates any linear precoding scheme, including zero-forcing (ZF) 
and MMSE precoding, without altering 
the sensing or tracking pipeline. CBF is adopted here to keep the 
notation compact within the page limit.}
\begin{equation}\label{eq:cbf}
    [\bW_c]_{:,k} = \sqrt{\eta_{k}}\,\bh^{*}_{k}, 
    \quad k \in \mathcal{K}_0,
\end{equation}
where $\eta_k \geq 0$ is the power coefficient for user $k$ 
at the CBS, subject to $\sum_{k\in\mathcal{K}_0}\eta_k
\|\bh_k\|^2 \leq N_b p_d$. The sensing probing vector is steered 
toward the current UAV angle estimate $(\hat{\theta}_t, \hat{\phi}_t)$:
\begin{equation}\label{eq:radar_beam}
    \bs_r = \sqrt{N_b p_d}\,\ba(\hat{\theta}_t,\hat{\phi}_t),
    \quad \|\bs_r\|^{2} = N_b p_d.
\end{equation}
For the first coherent interval, an acquisition sector is 
used in place of $(\hat{\theta}_t,\hat{\phi}_t)$; thereafter 
\eqref{eq:radar_beam} is updated from the tracking output. 

\subsubsection{Downlink Transmission and UAV Probing}

The CBS broadcasts $\bd$ per (\ref{eq:cbs_dl}). Each active AP $m \in \mathcal{M}_{\text{act}}$ simultaneously transmits:
\begin{equation}\label{eq:ap_tx}
s_m = \sqrt{p_d} \sum_{k \in \mathcal{K}_m} \sqrt{\eta_{mk}}\, \hat{g}^{*}_{mk}\, u_k,
\end{equation}
where $u_k$ is zero-mean unit-variance  data symbol for user $k$, and $\eta_{mk}$ denotes power coefficient for user $k$ at AP $m$.
Idle APs do not transmit. The received signal at UE~$k$ is
\begin{equation}\label{eq:rx_ue}
y_k = \mathbf{h}^{T}_{k} \bd + \sum_{m=1}^{M-N_b} g_{mk}\, s_m + n_k,
\end{equation}
where $n_k \sim \mathcal{CN}(0, \sigma^{2}_{n})$.

\subsubsection{Echo Reception}
Since the CBS has full knowledge of its own transmitted signal $\bd$, 
the communication component of the reflected echo can be cancelled 
prior to sensing processing. The residual monostatic echo after 
self-interference cancellation is
\begin{equation}\label{eq:mono_echo}
    \by_{b} = \sqrt{\rho}\,\xi_t\,e^{j2\pi f_{D,bt}t}\,
    \mathbf{h}_{bt}\mathbf{h}_{bt}^{T}\,\bs_r + \bn^{\text{echo}}_{b},
\end{equation}
with $\bn^{\text{echo}}_{b} \sim \mathcal{CN}(\mathbf{0},\sigma^{2}_{n}
\bI_{N_b})$. Similarly, the bistatic echo at AP~$m$, where $m \in \mathcal{M}_{\text{sense}} \triangleq 
\mathcal{M}_{\text{idle}} $ is
\begin{equation}\label{eq:bi_echo}
    r_m = \sqrt{\rho}\,\xi_t\,e^{j2\pi f_{D,mt}t}\,h_{mt}\,
    \mathbf{h}_{bt}^{T}\,\bs_r + n_m,
\end{equation}
with $n_m \sim \mathcal{CN}(0,\sigma^{2}_{n})$.
Each AP forwards $r_m$ to the CBS via fronthaul. The fronthaul overhead for sensing 
is one complex scalar per AP per symbol period, negligible relative 
to the $|\mathcal{K}_m|$-dimensional FU detection statistics 
forwarded by active APs.

\subsubsection{Parameter Estimation}
The CBS fuses $\by_{b}$ and 
$\{r_m : m \in \mathcal{M}_{\text{sense}}\}$. The CBS applies matched filtering 
with the known steering vector:
\begin{equation}\label{eq:mono_mf}
    z_b = \ba^{H}(\hat{\theta}_t,\hat{\phi}_t)\,\by_{b}.
\end{equation}
As $\mathbf{h}_{bt} =\sqrt{\zeta_{bt}}\,\ba(\theta_t,\phi_t)$:
\begin{equation}\label{eq:mono_mf_approx}
    z_b = \sqrt{\rho}\xi_t\,e^{j2\pi f_{D,bt}t}\,
    \zeta_{bt}\,N_b\sqrt{p_d} + \tilde{n}_b,
\end{equation}
with $\tilde{n}_b \sim \mathcal{CN}(0,\sigma^{2}_{n})$. The CBS weights the 
$M_s = |\mathcal{M}_{\text{sense}}|$ bistatic scalars by 
$\sqrt{\zeta_{mt}}\,e^{-j\psi_{mt}}$ (maximum-ratio combining 
over large-scale fading) and sums them:
\begin{equation}\label{eq:bi_mrc}
    z_{\text{bi}} = \sum_{m \in \mathcal{M}_{\text{sense}}} 
    \sqrt{\zeta_{mt}}\,e^{-j\psi_{mt}}\,r_m.
\end{equation}
The combined statistic $\mathbf{z} = [z_b,\,z_{\text{bi}}]^T$ is 
used for UAV angle, range, and Doppler estimation, following the 
generalised likelihood ratio test (GLRT) detection framework 
of~\cite{Ren2024}.
\section{Performance Analysis}
\label{sec:analysis}

\subsection{Uplink Spectral Efficiency}

The uplink SE derivation follows Section~IV of \cite{Jiang2024ICC} 
unchanged; $\rho$ does not affect uplink processing. For NU 
$k \in \mathcal{K}_0$, the CBS applies matched filtering with 
$\bh_k$ and, using the use-and-then-forget bounding technique of 
\cite{Ngo2017}, obtains:
\begin{equation}\label{eq:ul_nu}
    R^{\text{ul}}_{\text{nu},k} = \log_2(1 + \gamma^{\text{ul}}_{\text{nu},k}),
\end{equation}
\begin{equation}\label{eq:ul_nu_sinr}
    \gamma^{\text{ul}}_{\text{nu},k} = \frac{\eta_k N_b \alpha^{0}_{k}}
    {\sum_{j=1}^{K} \eta_j \beta^{0}_{j} - \eta_k \alpha^{0}_{k} 
    + \frac{\sigma^{2}_{n}}{p_u}}.
\end{equation}
For FU $k \notin \mathcal{K}_0$:
\begin{equation}\label{eq:ul_fu_sinr}
    \gamma^{\text{ul}}_{\text{fu},k} = 
    \frac{\eta_k \left(\sum_{m \in \mathcal{M}_k} \alpha_{mk}\right)^2}
    {\sum_{m \in \mathcal{M}_k} \alpha_{mk} \sum_{j=1}^{K} \eta_j \beta_{mj} 
    + \frac{\sigma^{2}_{n}}{p_u} \sum_{m \in \mathcal{M}_k} \alpha_{mk}}.
\end{equation}
The uplink sum SE is $C^{\text{ul}} = \sum_{k \in \mathcal{K}_0} 
R^{\text{ul}}_{\text{nu},k} + \sum_{k \notin \mathcal{K}_0} 
R^{\text{ul}}_{\text{fu},k}$.

\subsection{Downlink Spectral Efficiency}

In the downlink, each UE knows only channel statistics. 
Decomposing (\ref{eq:rx_ue}) for NU $k \in \mathcal{K}_0$ into 
desired signal, channel estimation error, channel uncertainty 
error, inter-user interference, sensing interference, 
and noise terms---following the use-and-then-forget bounding 
technique of \cite{Ngo2017,Elfiatoure2024}---yields:
\begin{equation}\label{eq:dl_nu}
    R^{\text{dl}}_{\text{nu},k} = \log_2(1 + \gamma^{\text{dl}}_{\text{nu},k}),
\end{equation}
\begin{equation}\label{eq:dl_nu_sinr}
    \gamma^{\text{dl}}_{\text{nu},k} = 
    \frac{\left(\sum_{n=1}^{N_b}\sqrt{(1-\rho)\eta_k}\,\alpha^{0}_{k}
    \right)^{2}}{D^{(\text{nu})}_{k} + I^{(s)}_{k}},
\end{equation}
where
\begin{align}\label{eq:dl_nu_denom}
    D^{(\text{nu})}_{k} &= \sum_{n=1}^{N_b}(1-\rho)\beta^{0}_{k}
    \sum_{j \in \mathcal{K}_0}\eta_j \alpha^{0}_{j} \notag\\
    &\quad + \sum_{m=1}^{M-N_b}\beta_{mk}\sum_{j \in \mathcal{K}_m}
    \eta_{mj}\alpha_{mj} + \frac{\sigma^{2}_{n}}{p_d},
\end{align}
\begin{equation}\label{eq:sensing_interf}
    I^{(s)}_{k} = \rho\, N_b p_d\,\big|\bh_k^{H}\ba(\hat{\theta}_t,
    \hat{\phi}_t)\big|^{2} \approx \rho\, p_d\, \beta^{0}_{k},
\end{equation}
where the approximation uses the massive MIMO near-orthogonality 
property $|\bh_k^{H}\ba|^{2} \approx \beta^{0}_{k}/N_b$ 
\cite{Demirhan2024}, so that the $N_b$ factors cancel. The sensing 
interference $I^{(s)}_{k}$ is proportional to $\rho$, making the 
communications--sensing tradeoff explicit at the receiver level.

For FU $k \notin \mathcal{K}_0$:
\begin{equation}\label{eq:dl_fu_sinr}
    \gamma^{\text{dl}}_{\text{fu},k} = 
    \frac{\left(\sum_{m \in \mathcal{M}_k}\sqrt{(1-\rho)\eta_{mk}}\,
    \alpha_{mk}\right)^{2}}{D^{(\text{fu})}_{k} + I^{(s)}_{k}},
\end{equation}
where $D^{(\text{fu})}_{k}$ is the FU interference-plus-noise 
denominator from \cite{Jiang2024ICC}. The downlink sum SE is 
$C^{\text{dl}} = \sum_{k \in \mathcal{K}_0} R^{\text{dl}}_{\text{nu},k} 
+ \sum_{k \notin \mathcal{K}_0} R^{\text{dl}}_{\text{fu},k}$.

\subsection{Multi-Static Sensing SNR}

Following the multi-static sensing SNR definition of 
\cite{Demirhan2024,Ren2024}, we define the total sensing SNR at 
the CBS as the ratio of expected signal power to noise power across 
all sensing observations. After matched filtering (\ref{eq:mono_mf}), we have
\begin{equation}\label{eq:snr_mono}
    \gamma^{\text{sense}}_{b} = 
    \frac{|\xi_t|^{2}\,\zeta_{bt}^{2}\,N^{2}_{b}\,\rho\, p_d}
    {\sigma^{2}_{n}}.
\end{equation}
This scales as $N^{2}_{b}$, reflecting the transmit array gain 
$N_b$ from the probing beam $\|\bs_r\|^2 = N_b p_d$ and the 
receive matched-filter gain $N_b$ of the co-located CBS aperture. After MRC combining (\ref{eq:bi_mrc}), each sensing AP 
$m \in \mathcal{M}_{\text{sense}}$ contributes:
\begin{equation}\label{eq:snr_bi}
    \gamma^{\text{sense}}_{m} = 
    \frac{|\xi_t|^{2}\,\zeta_{mt}\,\zeta_{bt}\,N_b\,\rho\, p_d}
    {\sigma^{2}_{n}}.
\end{equation}
This scales as $N_b$ (transmit array gain only), consistent with 
\cite{Demirhan2024}, since each AP receives with a single antenna.

Total fused sensing SNR is given by
\begin{equation}\label{eq:snr_total}
    \gamma^{\text{sense}} = \gamma^{\text{sense}}_{b} + 
    \sum_{m \in \mathcal{M}_{\text{sense}}} \gamma^{\text{sense}}_{m}.
\end{equation}
The monostatic term scales as $N^{2}_{b}$ while each bistatic term 
scales as $N_b$. Nevertheless, for large $|\mathcal{M}_{\text{sense}}|$
---as occurs under light user loading $K \ll M$---the aggregate 
bistatic sum may dominate. This confirms the unique value of the 
idle-AP receiver network, which has no analogue in CF-ISAC.

\subsection{Communications--Sensing Pareto Frontier}

Equations~(\ref{eq:dl_nu_sinr})--(\ref{eq:dl_fu_sinr}) show that 
the downlink SE of each user degrades monotonically in $\rho$, 
while (\ref{eq:snr_total}) shows that the sensing SNR grows 
monotonically in $\rho$. The Pareto frontier is therefore traced 
by sweeping $\rho \in [0,1]$ with all other system parameters fixed. A key quantitative observation: a small $\rho$ induces a linear sensing SNR gain ($\gamma^{\text{sense}} \propto \rho$) while 
causing a sub-linear SE loss, since the sensing interference 
$I^{(s)}_k \approx \rho\, p_d\,\beta^{0}_{k}$ is typically small 
relative to the desired signal power in the massive MIMO regime. 
This convexity of the SE curve in $\rho$ implies that the 
attractive operating regime is small $\rho$ ($\lesssim 0.2$), 
consistent with the numerical observations of 
\cite{Demirhan2024,Mao2024}.

\subsection{Fronthaul Overhead Analysis}

We extend the fronthaul comparison of \cite{Jiang2024ICC} to 
include the sensing echo backhaul. Per symbol period, the number 
of complex scalars exchanged over the fronthaul is:
\begin{itemize}
    \item \emph{CF}: $\sum_{k\in \mathbb{K}} |\mathcal{M}_k|$ symbols (with UC clustering $\mathcal{M}_k$).
    \item \emph{HCF}: $\sum_{k\notin \mathbb{K}_0} |\mathcal{M}_k|$ 
    symbols.
    \item \emph{HCF-ISAC}: $ |\mathcal{M}_{\text{sense}}|+\sum_{k\notin \mathbb{K}_0} |\mathcal{M}_k| $ symbols.
\end{itemize}
Since $|\mathcal{M}_{\text{sense}}| \leq M - N_b$ and each sensing 
echo is a single scalar (versus the multi-symbol FU detection 
statistics), the ISAC overhead increment over HCF communications 
is modest. HCF-ISAC thus achieves lower total fronthaul load than 
conventional CF while providing both communications service and 
sensing capability.

\section{Performance Evaluation}
\label{sec:results}

\subsection{Simulation Setup}

We consider a circular coverage area of radius $R = 1$\,km with a total antenna budget $M = 256$. The CBS, located at the centre, is equipped with $N_b = 128$ co-located antennas, while the remaining $128$ antennas are deployed as single-antenna APs uniformly distributed in the annular ring $[r, R]$. We set $K = 16$ single-antenna UEs placed uniformly in the disc. We compare HCF-ISAC against conventional CF-ISAC~\cite{Demirhan2024}, in which all $M = 256$ antennas are single-antenna APs---each contributing to both communications and sensing---without a co-located CBS. 
Communications channels follow the three-slope COST-Hata model~\cite{Ngo2017} with carrier frequency $f_c = 1.9$\,GHz, AP/CBS height $15$\,m, UE height $1.65$\,m, and log-normal shadowing with standard deviation $\sigma_{\text{sd}} = 8$\,dB. The noise power is computed as $\sigma^2_n = -174 + 10\log_{10}(B) + F$ in dBm, where bandwidth $B = 5$\,MHz and noise figure $F = 9$\,dB, yielding $\sigma^2_n \approx -98$\,dBm. Per-antenna downlink power is $p_d = 200$\,mW and UE uplink power $p_u = 100$\,mW. Equal power allocation is used at both the CBS and the APs. User-centric clustering at the APs follows a distance-based threshold: AP~$m$ serves FU~$k$ if $d_{mk} < 0.75\, \bar{d}_k$, where $\bar{d}_k$ is the mean distance from all APs to user~$k$. Results are averaged over $10^5$ independent random network realizations. For the sensing target, a UAV is placed at fixed altitude $H_t = 100$\,m with horizontal distance drawn uniformly from $r_t \in [200, 800]$\,m. The RCS is set to $\sigma_{\text{RCS}} = 1$\,m$^2$. 
\begin{figure}[t]
\centering
 \includegraphics[width=0.87\columnwidth]{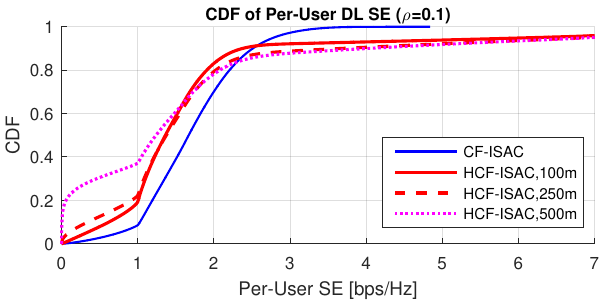}
\caption{CDFs of per-user downlink SE for HCF-ISAC with varying inner radius and CF-ISAC baseline. The power-splitting factor is $\rho = 0.1$.}
\label{fig:cdf_se}
\end{figure}
 
\begin{figure}[t]
\centering
\includegraphics[width=0.87\columnwidth]{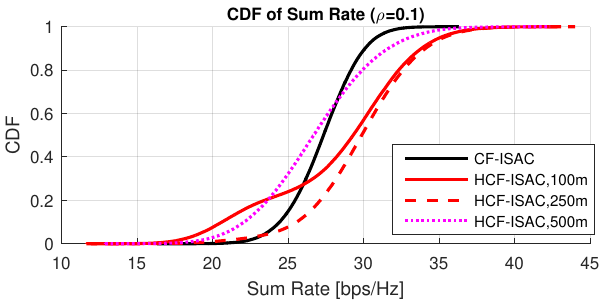}
\caption{CDFs of downlink sum rate for HCF-ISAC and CF-ISAC.}
\label{fig:cdf_sum}
\end{figure}
 \subsection{Numerical Results}

\begin{figure}[t]
\centering
 \includegraphics[width=0.87\columnwidth]{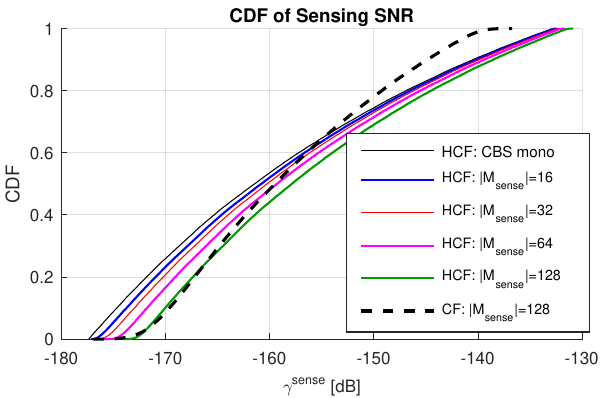}
\caption{CDF of multi-static sensing SNR $\gamma^{\text{sense}}$ versus the number of participating APs $|\mathcal{M}_{\text{sense}}|$, with $r = 100$\,m and $\rho = 0.1$. The CF-ISAC baseline with $|\mathcal{M}_{\text{sense}}| = 128$ is also shown.}
\label{fig:cdf_snr}
\end{figure}

Fig.~\ref{fig:cdf_se} shows the cumulative distribution functions (CDFs) of per-user downlink SE with $\rho = 0.1$ and $r \in \{100, 250, 500\}$\,m. Across the upper portion of the CDF, every HCF-ISAC curve lies to the right of CF-ISAC. This gain originates from the CBS's $N_b = 128$ co-located array, which provides coherent beamforming gain proportional to $N_b$ for the NUs---an advantage absent in CF-ISAC, where all $M = 256$ antennas are single-antenna APs. In the lower tail of the CDF (e.g., 5th percentile), the ordering reverses: CF-ISAC outperforms all HCF-ISAC variants. This is expected: CF-ISAC's fully distributed architecture ensures that every user has nearby APs. There is a deliberate architectural trade-off: HCF-ISAC sacrifices some user fairness in exchange for two structural benefits: (i)~the CBS array delivers substantially higher SE to NUs, raising the sum rate, as illustrated by Fig.~\ref{fig:cdf_sum}; and (ii)~only $M - N_b = 128$ APs require fronthaul connections---half the $M = 256$ links needed in CF-ISAC. This halving of the fronthaul network scale is a key practical advantage, as fronthaul cost and complexity are major deployment barriers for cell-free systems.

Fig.~\ref{fig:cdf_snr} plots the CDF of the fused sensing SNR $\gamma^{\text{sense}}$ defined in~(\ref{eq:snr_total}), evaluated for varying numbers of participating APs $|\mathcal{M}_{\text{sense}}|$. We set $r = 100$\,m and $\rho = 0.1$. Even without any AP participation ($|\mathcal{M}_{\text{sense}}| = 0$, CBS monostatic only), HCF-ISAC achieves a non-trivial sensing SNR driven by the $N_b^2$ monostatic term in~(\ref{eq:snr_mono}). The co-located CBS array provides both transmit beamforming gain ($N_b$) and coherent receive combining gain ($N_b$). As $|\mathcal{M}_{\text{sense}}|$ increases from $16$ to $128$, the CDF shifts progressively to the right. Each doubling of $|\mathcal{M}_{\text{sense}}|$ provides an approximately $3$\,dB gain, consistent with the linear accumulation of bistatic SNR contributions in~(\ref{eq:snr_total}). The CF-ISAC baseline with $|\mathcal{M}_{\text{sense}}| = 128$ single-antenna APs is also shown. Despite having the same number of receivers, CF-ISAC achieves lower sensing SNR than HCF-ISAC. This is because CF-ISAC lacks the powerful $N_b^2$ monostatic contribution from a co-located array. The results show that even a modest number of idle APs (e.g., $|\mathcal{M}_{\text{sense}}| = 32$) combined with the CBS monostatic echo can substantially improve sensing coverage. In practical deployments where not all APs are actively serving users at every time slot, HCF-ISAC can opportunistically exploit these idle APs for sensing without any additional infrastructure cost.

\begin{figure}[t]
\centering
\includegraphics[width=0.87\columnwidth]{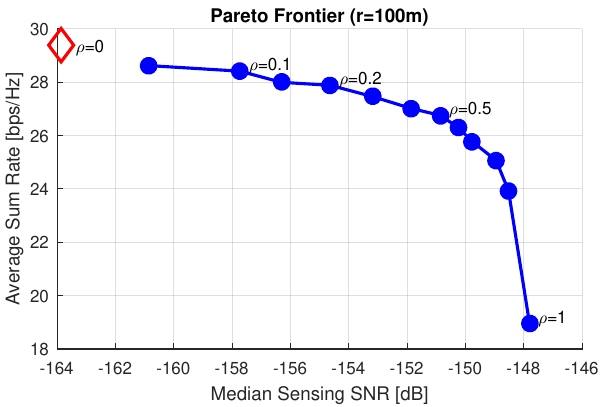}
\caption{Communications--sensing Pareto frontier.}
\label{fig:pareto}
\end{figure}

Fig.~\ref{fig:pareto} shows the Pareto curve of average sum rate versus median sensing SNR as the power-splitting factor $\rho$ varies from $0$ to $1$. For $\rho \in [0, 0.2]$, the sum rate decreases only mildly while the sensing SNR improves substantially. This gentle slope reflects the fact that only the NUs are affected by $\rho$, and the $(1 - \rho)$ scaling of their signal power causes a logarithmic (hence slowly varying) SE reduction. Beyond $\rho \approx 0.3$, the sum rate drops more steeply. At $\rho = 0.5$, half of the CBS power is devoted to sensing, reducing NU SEs significantly. At $\rho = 1$ (all CBS power to sensing, no communications from the CBS), only the FUs contribute to the sum rate via the APs. The ``knee'' of the Pareto curve lies at $\rho \in [0.1, 0.2]$, where the system achieves strong sensing capability with only a modest communications cost. This operating region is consistent with the communications--sensing trade-off findings in~\cite{Mao2024} and confirms that HCF-ISAC can deliver effective dual-function performance with a single, easily tunable parameter.
 
\section{Conclusions}
\label{sec:conclusion}
We proposed HCF-ISAC massive MIMO, the first integration of ISAC into the hierarchical cell-free and cellular architecture. For a UAV target, we defined precise node roles, a five-phase ISAC protocol, and closed-form SE and sensing SNR expressions. The sensing SNR decomposes into a monostatic term scaling as $N^{2}_{b}$ and a bistatic sum scaling linearly with the number of idle APs. Numerical results confirm that HCF-ISAC simultaneously achieves higher sum throughput than CF-ISAC, superior sensing SNR, under lower fronthaul costs. Future work includes joint beamformer optimization, multi-target extension, rigorous characterisation of pilot contamination from UAV reflections, and sensing-privacy analysis of the CBS probing beam.

\end{document}